\documentclass{emulateapj}

\newcommand{\ngc}{NGC\,1097}

\newcommand{\qm}[1]
{``#1''}

\begin{document}

\title{RADIATIVELY INEFFICIENT ACCRETION FLOW IN THE NUCLEUS OF \ngc}

\author{RODRIGO S. NEMMEN and THAISA STORCHI-BERGMANN}
\affil{Instituto de F\'isica, Universidade Federal do Rio Grande do Sul,
Campus do Vale, Porto Alegre, RS, Brasil}
\email{rodrigo.nemmen@ufrgs.br}

\author{FENG YUAN}
\affil{Shanghai Astronomical Observatory,
Chinese Academy of Sciences, 80 Nandan Road, Shanghai 200030, China}

\author{MICHAEL ERACLEOUS}
\affil{Department of Astronomy and
Astrophysics, The Pennsylvania State University, USA}

\author{YUICHI TERASHIMA}
\affil{Institute of Space and Astronautical Science, Japan}
\and
\author{ANDREW S. WILSON}
\affil{Astronomy Department, University of Maryland, USA}

\begin{abstract}
We present a model for the accretion flow around the supermassive black hole in
the LINER nucleus of \ngc\ which fits the optical to X-ray spectral
energy distribution (SED). The X-ray segment of the SED is based on
observations with the {\it Chandra X-Ray Observatory}, which are
reported here for the first time.  The inner part of the flow is
modeled as a radiatively inefficient accretion flow (RIAF) and the
outer part as a standard thin disk.  The value of the transition
radius ($r_{tr} \approx 225\,R_S$, where $R_S=2GM/c^2$) between the RIAF and
outer thin disk was obtained from our previous fitting of the
double-peaked Balmer emission line profile, which originates in the
thin disk. The black hole mass was inferred from measurements of the
stellar velocity dispersion in the host galaxy.  When these parameters
are used in the accretion flow model, the SED can be successfully
reproduced, which shows that the line profile model and the accretion
flow model are consistent with each other.  A small remaining excess
in the near-UV is accounted by the contribution of an obscured
starburst located within 9~pc from the nucleus, as we reported
in an earlier paper.  The radio flux is consistent with synchrotron
emission of a relativistic jet modeled by means of the internal shock
scenario.  In an appendix we also analyze the {\it Chandra} X-ray
observations of the $\sim$ 1 kpc circumnuclear star-forming ring and of an
ultraluminous compact X-ray source located outside the ring.
\end{abstract}

\keywords{accretion, accretion disks --- black hole physics ---
galaxies: individual (\ngc) --- galaxies: active --- galaxies: nuclei}

\section{INTRODUCTION} \label{sec:intro}

Active galactic nuclei (hereafter AGNs) are thought to be
powered by accretion onto a supermassive black hole. The accreting
matter is thought to form a thin, equatorial accretion disk, which may
reveal itself observationally via its continuum emission (the \qm{big
blue bump}, produced by the sum of black body spectra from annuli in
the disk at different temperatures; e.g., \citealt{frank02,koratkar99}).
Another possible signature of a thin accretion disk is the profile of
the Fe~K$\alpha$ X-ray line, which is emitted from the inner parts of
the disk and distorted by strong relativistic effects \citep[see, for
example,][]{fabian89,tanaka95}. An alternative signature, broad,
double-peaked Balmer emission lines from the outer accretion disk
(analogous to the lines of cataclysmic variables) has been observed in
the optical spectra of some AGNs \citep[see review
by][]{eracleous98}.

Many AGNs with double-peaked Balmer lines (hereafter
``double-peakers'') were discovered in the spectroscopic survey of
radio-loud AGNs \citep{eh94,eh03} and by the Sloan Digital Sky Survey
\citep{strateva03,wang05}.  However, \ngc\ is an important object
because it was the first (and the best studied yet) case of a
low-luminosity LINER nucleus to display broad, double-peaked H$\alpha$
and H$\beta$ emission lines \citep{sb93}, with a full width at half
maximum of $\approx 7500$~km~s$^{-1}$. This discovery was followed by
observations of such lines in other low-luminosity AGNs (LLAGNs),
mostly with the \textit{Hubble Space Telescope} (HST,
\citealt{bower96,shields00,ho00,barth01}).
The appearance of the disklike emission lines of \ngc\ is thought to
be a transient event because they were not seen in previous
observations. This transient event was possibly a tidal disruption of
a star by a nuclear supermassive black hole, where the debris of the
disrupted star feeds the accretion flow, although other
possibilities were also discussed by \citet{sb97}. Since \ngc\ is a
LLAGN, this transient phenomenon opened up the possibility that we
could witness variations in the disklike emission lines on a timescale
of years or even months, and the variability of the broad H$\alpha$
lines was indeed monitored (\citealt{sb95,sb03}, hereafter SB03). The
observed H$\alpha$ profiles of \ngc\ were reproduced by kinematical
models of non-axisymmetric thin accretion disks by \citet{sb97}, who
inferred that the line-emitting disk is truncated at an inner radius
of $\approx 225 R_S$, where $R_S=2 G M / c^2$ is the Schwarzschild
radius and $M$ is the black hole mass, the black hole mass was measured by \citet{lewis05} as $(1.2 \pm 0.2) \times 10^8 M_\odot$ from the stellar velocity dispersion using the Ca~II near-IR triplet lines.

On the assumption that the broad double-peaked emission lines observed
in AGNs originate in a thin accretion disk, an external source of
illumination is needed to power them. Some compelling arguments for
this have been put forward by different authors, involving the energy
budget of the line-emitting region (e.g., \citealt{chen89a}) and the
low temperatures achieved by the disk under local viscous dissipation
\citep{collin87}.  For instance, considering the energy budget of the
thin disk in a small sample of {\it radio-loud} double-peakers
\citep{eh94,eh03}, the disk can power its own lines only if virtually
all the energy available locally is converted into recombination
lines, which is unlikely (\citealt{chen89a}).  The presence of an
inner, inflated, radiatively inefficient accretion flow (RIAF, see
reviews by \citealt{quataert01,narayan98r}, hereafter NMQ98) emitting
an X-ray continuum which illuminates the outer thin disk solves all
these energetic problems, as was proposed initially by \citet{chen89b}
in the form of the ion torus \citep[an early version of current RIAF
models; see][]{rees82}.

While spectral models of RIAFs have successfully accounted for the
spectral energy distributions (hereafter SEDs) of a handful of LINERs
(\citealt{lasota96,quataert99,ptak04}, but see
\citealt{yuan02b,herrnstein05} in the case of NGC 4258), RIAFs have
yet to be studied in a double-peaker LLAGN. The reason for this is
that there was no object until now which had both good observations of
the nuclear SED and displayed disklike broad emission lines. We have
now such observations for \ngc, which have allowed us to investigate
the structure of the accretion flow in the nucleus by comparing the
observed broadband SED with predictions from theoretical models in
which the plasma at small radii is in the form of a RIAF and the outer
parts of the flow are a standard thin disk. We explore the conditions
under which the parameters of the thin disk inferred from the
H$\alpha$ profile model presented by SB03 are consistent with the
modelling of the continuum. In \textsection \, \ref{sec:obs} we
describe the multiwavelength data comprising the SED, including new
X-ray observations as well as previously published data in the
optical/ultraviolet, infrared and radio wavebands. In \textsection \,
\ref{sec:sed} we derive several properties of the SED, such as the
degree of radio-loudness and the bolometric and X-ray luminosities,
and we compare the SED of \ngc\ with the typical SEDs of quasars and
LLAGNs. In \textsection \, \ref{sec:riaf} we discuss the accretion
flow model and compare its predictions with the observations. We
discuss our results in \textsection \, \ref{sec:discuss} and present
our conclusions in \textsection \, \ref{sec:conc}. Throughout this
paper we assume a distance\footnote{We note that \citet{sb05} assumed
a distance to \ngc\ of 17 Mpc. We choose the distance 14.5 Mpc in
order to compare directly our results for the SED with those of
\citet{ho00}, who adopt the latter distance to \ngc. Furthermore, this
difference in the assumed distance does not change our conclusions and
its effect is within flux calibration uncertainties.} to \ngc\ of
14.5~Mpc, thus $1\arcsec = 70$~pc. In an appendix, we present a
spectral analysis of the $\sim$ 1 kpc starforming ring and of an
ultraluminous compact X-ray source located outside the ring.

\section{OBSERVATIONS} \label{sec:obs}

In order to construct the SED of the nucleus of \ngc, we have used
X-ray observations from the \textit{Chandra X-ray Observatory} and
near-UV and optical observations from the \textit{Hubble Space
Telescope} (HST). We have also collected infrared and radio fluxes
from the literature. The sections below describe the data in detail.

\subsection{X-RAY DATA} \label{sec:x-ray}

\ngc\ was observed with the backside illuminated ACIS-S3 CCD aboard
the \textit{Chandra X-ray Observatory} on June 28, 2001. Table
\ref{tab:chandra-log} gives a log of the {\it Chandra} observations.

\begin{figure}[!ht]
\centering
\includegraphics[scale=0.55]{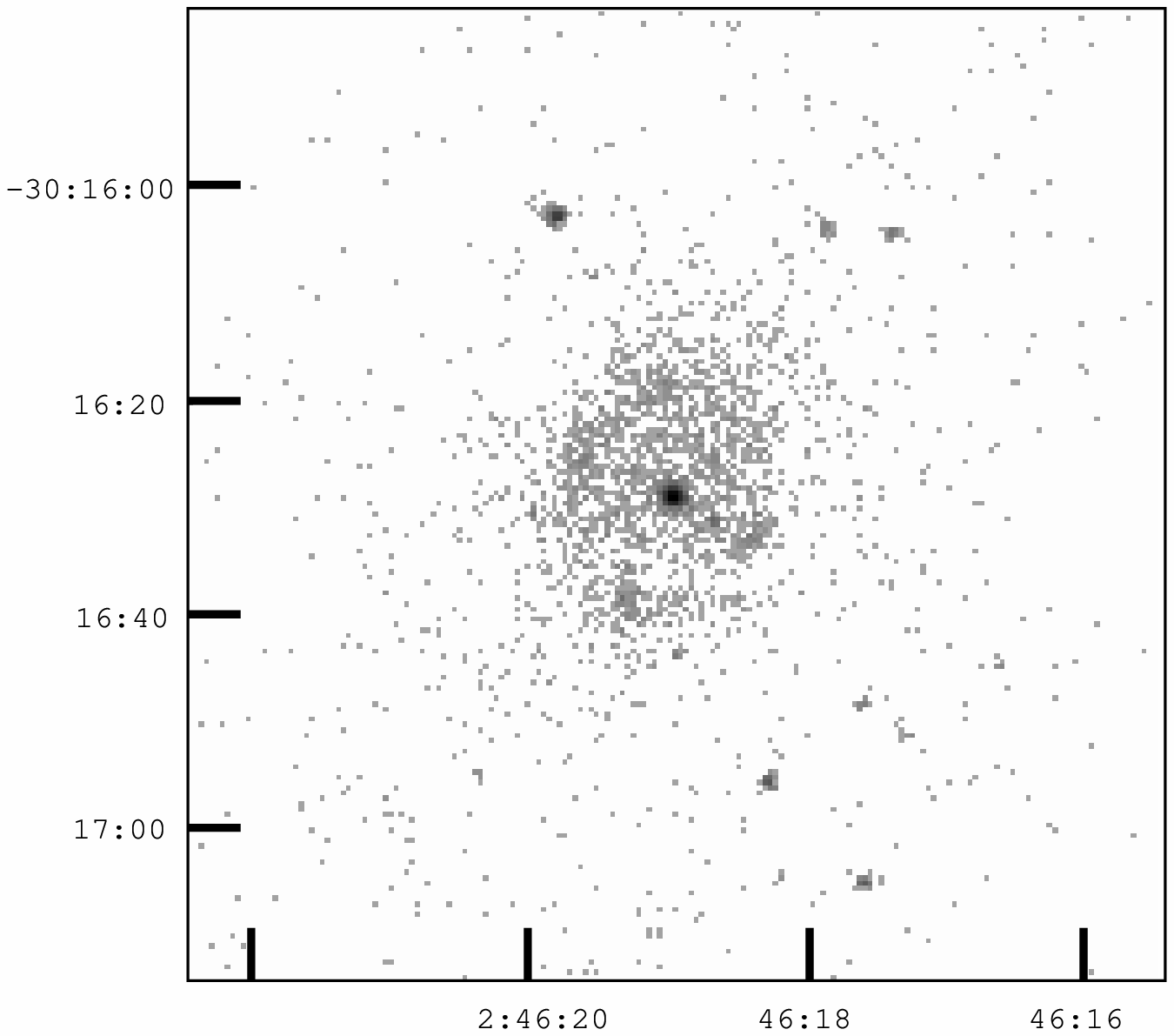}
\caption{X-ray image of the nucleus of \ngc\ (0.5 - 8 keV). The angular
size of the region shown corresponds to $90\arcsec \times 90\arcsec$ and the linear size is 6.3 kpc.  The ULX coordinates are 02h 46m 19.81s $-30^\circ$ 16' $02\farcs8$ (J2000).}
\label{fig:chandra-image}
\end{figure}

Figure \ref{fig:chandra-image} shows the image of the central region
of the galaxy in the 0.5--8.0~keV band, which illustrates the extended
emission morphology. The nucleus (black spot in the center) and
circumnuclear region (including the star-forming ring at $\sim
9\arcsec$) are clearly seen in the image. An ultraluminous compact
X-ray source (ULX) is observed at $27\farcs6$ NNE from the
nucleus. The properties of the star-forming ring and ULX are discussed in
the Appendix.

An X-ray spectrum of the nucleus was extracted from a circular region
of radius $2\farcs6$. The effect of pile up is too large in the first
observation to measure reliable spectral shape and flux. The second
observation is also affected by pile up, but the effect was
relatively small because the detector was operated in 1/8 sub-array
mode. We accounted for pile-up by applying the pile-up model
implemented in the
XSPEC\footnote{http://heasarc.gsfc.nasa.gov/docs/xanadu/xspec}
spectral-fitting package.

\begin{figure}[!ht]
\centering
\includegraphics[scale=0.35,angle=-90]{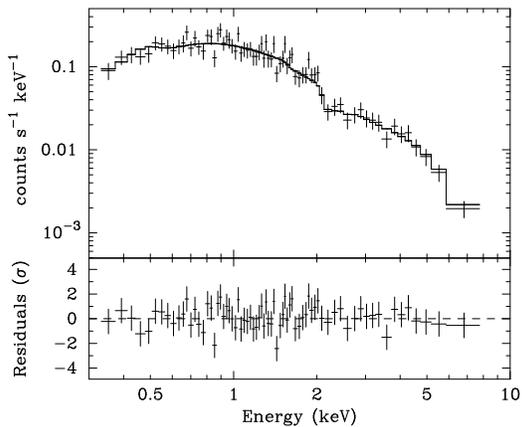}
\caption{X-ray spectrum of \ngc. The upper panel shows the data and the
best-fit model, and the lower panel shows the residuals of the fit.}
\label{fig:chandra-nuc}
\end{figure}

Spectral fits were performed to the X-ray spectrum, using a $\chi^2$ minimization technique.
Acceptable fits were obtained using a power-law model ($I_E \propto E^{-\Gamma}$, where $I_E$ is a photon flux in units of ${\rm photons} \; {\rm s}^{-1} \: \rm{cm}^{-2} \: \rm{keV}^{-1}$
and $\Gamma$ is the photon index) modified
by photoelectric absorption along the line of sight. Figure
\ref{fig:chandra-nuc} shows the observed spectrum of the nucleus with
the best-fit model superposed. The model parameters for this fit are
listed in Table~\ref{tab:chandra-fits}, where $N_{\rm H}$ is the
hydrogen column density and $F_{2-10 \: \rm{keV}}$ is the observed
flux in the 2--10~keV band corrected for absorption.  The value of the
photon index is consistent with the typical values observed in LLAGNs
($\Gamma = 1.6-2.0$; see, e.g.,\citealt{terashima02,terashima03}).
The X-ray spectrum of the nucleus does not have a high enough signal-to-noise ratio to constrain spectral features such as Fe K emission.

The ASCA observation \citep{iyomoto96,terashima02} includes all of the
nucleus, the starforming ring and the ULX due to its poor angular
resolution. The hard component detected with ASCA is most likely
dominated by the nucleus, because the number of 2--8~keV photons from
a few faint off-nuclear sources and the weak hard emission from the
starforming ring is about 10\% that of the nucleus, thus the ASCA
spectral fits are reliable measures of the spectral shape of the
nucleus. The hard band flux obtained with ASCA ($2.07 \times 10^{-12}
\: {\rm erg} \; {\rm s}^{-1} \: \rm{cm}^{-2}$ in the 2--10~keV band or
$1.71 \times 10^{-12} \: {\rm erg} \; {\rm s}^{-1} \: \rm{cm}^{-2}$ in
the 2--8~keV band, corrected for absorption) is consistent with the
{\it Chandra} nuclear flux if the ASCA calibration uncertainty (10 \%)
and a small contribution from the off-nuclear hard X-ray emission are
taken into account.

\subsection{OPTICAL/UV DATA} \label{sec:hst}

The nucleus of \ngc\ was observed with the \textit{Space Telescope
Imaging Spectrograph} (STIS) in February 2001 in the wavelength range
1000~\AA--1~$\mu$m, using a $0\farcs2$ wide and $52\arcsec$ long slit.
The details of these observations are described in \citet{sb05}
(hereafter SB05), who have used these data to infer the presence of a
starburst within 9~pc from the nucleus.  The nuclear spectrum used in
the present work was extracted using a window of $0\farcs2 \times
0\farcs6$, the latter corresponding to 12 pixels along the slit.
Figure 1 of SB05 shows the nuclear spectrum, where the characteristic
broad double-peaked emission lines can be seen.

In order to obtain the intrinsic optical-to-UV (OUV) nuclear
continuum, we have to correct the spectrum for the effects of dust
extinction.  The Galactic extinction affecting the HST spectrum is
$E(B-V)=0.027 \; \rm{mag}$ \citep{schlegel98}. On the other hand, the
neutral hydrogen column density measured from the X-ray spectral
fit is $N_{\rm H} = 2.3 \times 10^{20} \; \textrm{cm}^{-2}$
(\textsection \, \ref{sec:x-ray}). Using the relation $E(B-V)=N_{\rm
H}/(5.8 \times 10^{21} \; \rm{cm}^{-2})$ mag \citep{bohlin78} we
obtain $E(B-V)=0.04$ mag. Subtracting the contribution of the Galactic
reddening above, the resulting internal reddening is $E(B-V)=0.013$
mag, which is very small. Adopting $R_V=A_V/E(B-V)=3.1$ the total
reddening affecting the observations is $A_{V}=0.124$ mag, which
implies only small extinction corrections to the spectrum using the
Galactic reddening law.

\subsection{INFRARED AND RADIO DATA} \label{sec:ir-radio}

Data in the infrared (IR) and radio were collected from the literature
and are listed in Table \ref{tab:ir-radio}.  \ngc\ has a luminous
circumnuclear ring of star formation of radius $\approx 9\arcsec$
\citep{sb96}. Most data available in the literature were obtained
using an aperture which includes this circumnuclear ring, and the
fluxes so measured can only be considered as upper limits to the
nuclear emission.
Therefore we decided not to include such observations in the SED, including those made with the IRAS and ISO satellites (apertures in the range $24\arcsec - 1\arcmin$). The only data
points we include in the SED are those corresponding to apertures smaller
than $10\arcsec$.

\section{THE SPECTRAL ENERGY DISTRIBUTION} \label{sec:sed}

\begin{figure*}[!ht]
\centering
\includegraphics[scale=0.9]{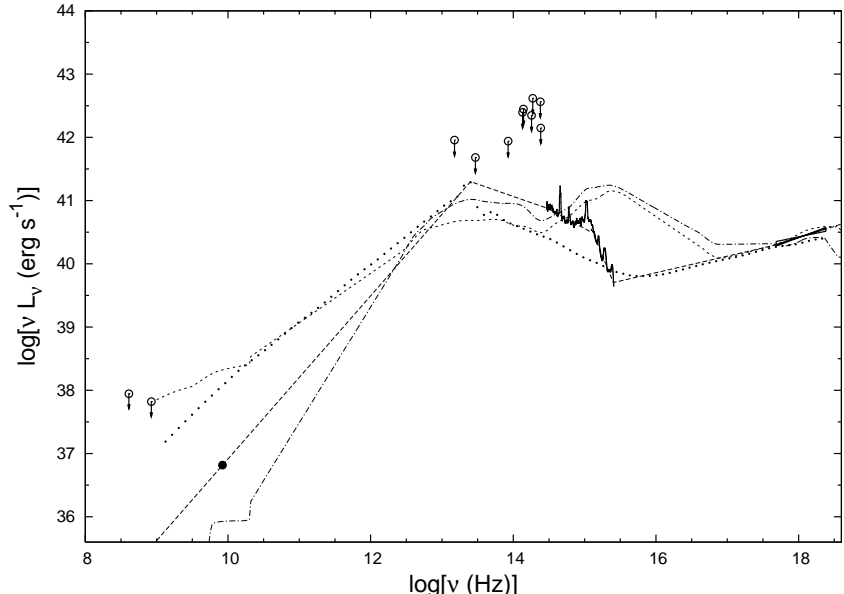}
\caption{Spectral energy distribution of the nucleus of \ngc\ and
power-law representations of the SED (\textit{long-dashed line}), as
described in the text. For comparison, the average SED of LLAGNs
(\textit{dotted line}, \citealt{ho05}) and the average SEDs of
radio-loud (\textit{short-dashed line}) and radio-quiet
(\textit{dot-dashed line}) quasars \citep{elvis94} are shown. The
spectra are normalized at \ngc's nuclear luminosity at 2 keV. The open
circles denote data possibly contaminated by the circumnuclear
starburst ring or host galaxy.}
\label{fig:sed}
\end{figure*}

We show in Figure \ref{fig:sed} the complete SED obtained by combining
the {\it Chandra}, HST, infrared and radio data obtained through
apertures smaller than $10\arcsec$. We notice that there
is a discontinuity between the IR and optical data. We believe this is due to additional components not included in our HST data (e.g. hot dust), and thus these points are treated as upper limits to the nuclear emission and are represented by open circles. The filled circle is a radio
data point observed with a small beam ($0\farcs25$).

In order to estimate the various quantities which characterize the
SED, we must deal with regions of the spectrum for which we have no
data available. Thus, we interpolate the SED with six power-laws of
the form $F_\nu \propto \nu^{-\alpha}$, one for each relevant
frequency interval, as shown by the long-dashed line in Figure
\ref{fig:sed}. The radio to infrared power-law was obtained by connecting
the radio luminosity value at 8.4~GHz to the average value of the normalized LLAGN SED at $2.5 \times 10^{13}$ Hz. The infrared to optical power-law was obtained by connecting the former point to the optical luminosity value at 1~$\mu$m. The OUV power-laws were obtained by fitting the data using least squares. The  power-law between $2.5 \times 10^{15}$ Hz and 2 keV was obtained by connecting the former point to the latter one. The best-fit X-ray power-law is used between 2 keV and 100 keV, above 100 keV the SED is described by an exponential cutoff.

\subsection{Properties of the SED} \label{sec:sed:props}

We use the interpolated piece-wise power-law described in the previous section to
calculate several properties of the SED: the radio loudness, bolometric
luminosity, X-ray luminosity and various useful spectral indices, as
follows.

\noindent
{\it Radio loudness --} The radio loudness can be quantified using two
different parameters: an optical to radio ratio defined as $R_o \equiv
L_\nu (6 \; {\rm cm}) / L_\nu ({\rm B})$ \citep{kellermann89} and a
radio to X-ray ratio defined as $R_{\rm x} \equiv \nu L_\nu (6 \; {\rm
cm}) / L_{\rm x}$, where $L_{\rm x}$ is the luminosity in the 2 -- 10
keV band \citep{terashima03}. The boundary between radio-loud and
radio-quiet objects corresponds to $R_o = 10$ and $R_{\rm x} = 3.162
\times 10^{-5}$. \ngc\ is marginally radio-loud according to
both criteria ($R_o = 11$, $R_{\rm x} = 7.6 \times 10^{-5}$), much
like other LLAGNs previously studied
\citep{ho99,ho01,ho02a,terashima03}. \citet{ho00} have quoted a value
of $R_o=19$, indicating a somewhat larger degree of radio-loudness,
but they do not specify what data they used to derive this value.

\noindent
{\it Bolometric luminosity --} The bolometric luminosity is $L_{\rm
Bol} = 8.6 \times 10^{41} \; \rm{erg} \; \rm{s}^{-1}$.  This value is
well within the range observed for other LLAGNs ($2.1 \times 10^{41} -
8.0 \times 10^{42} \; \rm{erg} \; \rm{s}^{-1}$, from
\citealt{ho99}). Adopting a black hole mass of $1.2 \times 10^8 M_\odot$ \citep{lewis05}, we obtain an Eddington ratio of $L_{\rm Bol} /
L_{\rm Edd} = 5.7 \times 10^{-5}$, where $L_{\rm
Edd} = 1.25 \times 10^{38} \: m = 1.5 \times
10^{46} \; \rm{erg} \; \rm{s}^{-1}$, (where $m$ is the black hole mass
in units of $M_{\odot}$).  For comparison, \citet{ho99} quotes a value
for this ratio in the interval $\sim 10^{-6} - 10^{-3}$ for
LLAGNs. $L_{\rm Bol}$ and $L_{\rm Bol}
/ L_{\rm Edd}$ are approximately two times smaller than the
values quoted by \citet{ho00}.

\noindent
{\it X-ray luminosity --} The
X-ray luminosity in the 2 -- 10 keV band is $L_{\rm x} = 4.4 \times
10^{40} \; \rm{erg} \; \rm{s}^{-1}$, which is in the interval of X-ray
luminosities of LLAGNs studied by \citet{terashima03}, $5 \times
10^{38} - 8 \times 10^{41} \; \rm{erg} \; \rm{s}^{-1}$.  The ratio of
$L_{\rm x}$ to narrow H$\alpha$ luminosity is 300, where the narrow
H$\alpha$ luminosity is $1.46 \times 10^{38} \; \rm{erg} \;
\rm{s}^{-1}$, measured from the HST spectrum. This ratio somewhat
exceeds the range of values estimated by \citet{terashima03}, 0.27 --
104. The ratio of X-ray luminosity in the 0.5 -- 10 keV band to
$L_{\rm Bol}$ is 0.08, which is within the range of values found by
\citet{ho99} for LLAGNs, 0.06 -- 0.33.

\noindent
{\it Spectral indices --} The value of the energy index of the
continuum in the hard X-ray band (2--10 keV) can be obtained from the
photon index (\textsection \, \ref{sec:x-ray}) using the relation
$\alpha_{\rm x} = \Gamma -1 = 0.64$. We measured the value of the
power-law index $\alpha_{\rm ou}$ of the OUV continuum in \textsection
\, \ref{sec:hst} as being 1.87.  The two-point spectral index
$\alpha_{\rm ox}$ from 2500 \AA -- 2 keV is 1.14, which is similar to
the mean value found by \citet{ho99} for LLAGNs, $\langle \alpha_{\rm
ox} \rangle \approx 0.9$.

\subsection{Comparison with SEDs of other AGNs} \label{sec:sed:comp}

Figure \ref{fig:sed} shows the SED of \ngc\ together with the average
SED of LLAGNs from \citet{ho05} and average SEDs of radio-loud and
radio-quiet quasars from \citet{elvis94}. All curves are normalized at
the 2~keV luminosity of \ngc.

In Figure \ref{fig:sed} it can be seen that in spite of \ngc's SED
being more radio-quiet than the average SED of LLAGNs, the general
shape of \ngc 's SED is similar to this mean SED. This result has been
verified quantitatively in the previous section through the similarity
between the values of the spectral indices from the two SEDs. It can
also be seen in Figure \ref{fig:sed} that the continuum of \ngc\ lacks
the canonical UV bump commonly present in quasars, which is associated
with emission from a geometrically thin, optically thick accretion
disk \citep{frank02}. This was already noted in the particular case of
\ngc\ by \citet{ho00} and more generally in LLAGNs by \citet{ho99},
and indicates an absence or truncation of the inner thin accretion
disk in these objects. The latter interpretation will be favored in
the modelling of the next section.

\section{MODELS FOR THE SED} \label{sec:riaf}

The observational characteristics of \ngc\ described in Section
\textsection \, \ref{sec:sed} (and in general of LLAGNs; see
\citealt{ho05} for a review) can be understood within the theoretical
framework of RIAFs (see reviews by \citealt{quataert01,narayan05};
early versions of RIAF models were called ADAFs or \qm{ion tori}; see
NMQ98, \citealt{kato98} for reviews). RIAFs are optically thin,
geometrically thick accretion flows that have radiative efficiencies
much less than the canonical 10\% of thin disks and thus naturally
generate low luminosities.  RIAFs are thought to arise when the mass
accretion rate is below a critical value $\dot{M}_{\rm crit} \approx
0.01 \dot{M}_{\rm Edd}$ (NMQ98). A low Eddington ratio is a
requirement for the existence of RIAFs ($L_{\rm Bol} / L_{\rm Edd} \la
0.01-0.1$; see, e.g., NMQ98) and the low ionizing luminosities
produced by RIAFs coupled with the typical densities of the narrow
line region lead naturally to low values of the ionization parameter
$U$, which consequently give rise to LINER-like spectra
\citep{halpern83,ferland83}.

In view of these indications, the model we adopt consists of an inner
RIAF plus an outer standard thin disk, which has been successful in
reproducing the SEDs of different kinds of sources, including black
hole X-ray binaries, Sgr A*, LLAGNs in elliptical galaxies and LINERs
(for reviews, see \citealt{quataert01,narayan05}).  We incorporate in
the modelling of the RIAF recent theoretical advances. As already
predicted in the pioneering papers \citet{narayan94,narayan95a}, and
confirmed by recent numerical simulations
\citep{stone99,hawley02,igumenshchev03,proga03} and analytical
calculations \citep{blandford99,narayan00,quataert00}, only a small
fraction of the gas supplied at large distances from the RIAF actually
falls on to the black hole, the rest of the gas is either ejected in
an outflow (advection-dominated inflow/outflow solution [\qm{ADIOS}])
or is prevented from accreting by convective motions
(convection-dominated accretion flow [CDAF]).

Given the theoretical uncertainties on the structure and microphysics
of the flow, we follow \citet{blandford99} and assume that the mass
accretion rate varies with radius as $\dot{M}(r) = \dot{M}_0 \left(
r/r_{\rm tr} \right)^p$, where $\dot{M}_0$ is the accretion rate at
the transition radius $r_{\rm tr}$ between the inner RIAF and the
outer thin disk. The exponent $p$ and the fraction of the turbulent
energy that heats the electrons, $\delta$, are free parameters in our
model.
This assumption has been recently confirmed by MHD simulations (e.g., \citealt{proga03}).

We assume $m=1.2 \times 10^8 M_\odot$ \citep{lewis05}, $r_{\rm out}=10^5$, where $r_{\rm
out}$ is the outer radius of the thin disk, and $i=34^{\circ}$ (SB03)
is the inclination of the axis of the disk to the line of sight. We set the plasma
parameter (ratio of gas to total pressure) to the value $\beta=0.9$
(for motivation, see \citealt{quataert99beta}) and we determine the
value of the viscosity parameter $\alpha$ from $\beta$ using the
equation $\alpha \approx (3/2) (1-\beta)/(3-2\beta) \approx 0.1$ (NMQ98,
\citealt{hawley96}).  We adjust the accretion rate to reproduce the
observed X-ray flux, and hereafter we use the accretion rate in
Eddington units $\dot{m}$ ($\dot{m}=\dot{M}/\dot{M}_{\rm Edd}$).  We
solve the radiative transfer and hydrodynamical equations of the RIAF
self-consistently, obtaining the emergent spectrum (see
\citealt{yuan03} for details). The radiative processes considered in
the RIAF are synchrotron and bremsstrahlung emission, and
Comptonization of these photons and the soft photons from the outer
thin disk. The thin disk is assumed to emit locally as a blackbody and
reprocesses X-ray photons from the RIAF.

The value of the transition radius was obtained from fitting the
double-peaked H$\alpha$ line profile with disk models as $\approx 225
R_S$ (SB03). On the other hand, there is also an independent constraint
on $r_{\rm tr}$ from fitting the SED with the accretion flow model,
i.e. the value of $r_{\rm tr}$ must be suitable to fit the OUV
spectrum. We set the value of $r_{\rm tr}$ to be that obtained from the
line profile fitting, i.e. $r_{\rm tr}= 225$.

\begin{figure*}[!ht]
\centering
\includegraphics[scale=0.9]{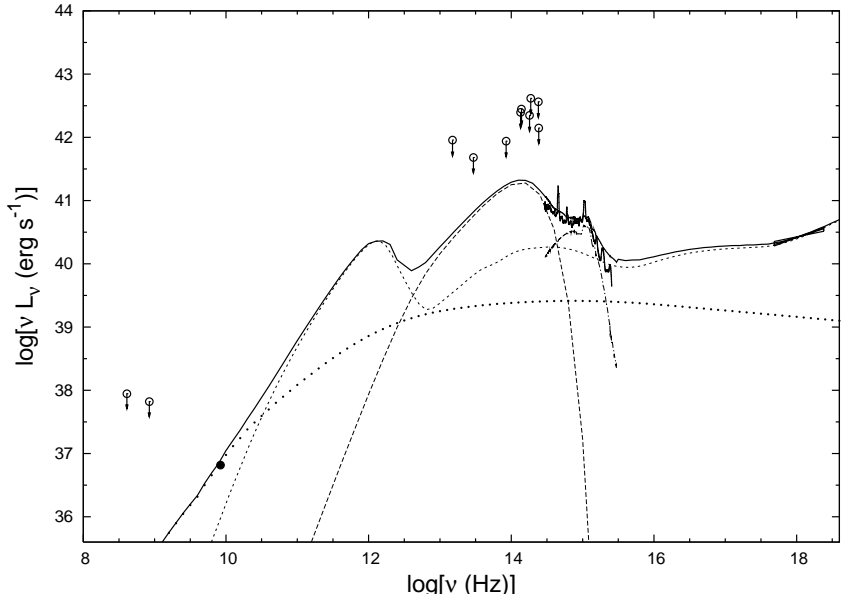}
\caption{Models of the RIAF (\textit{short-dashed line}), thin disk
(\textit{long-dashed line}), jet (\textit{dotted line}) and obscured
starburst (\textit{dot-dashed line}) compared to the nuclear SED of
\ngc. The sum of all components is also shown (\textit{solid line}). The thin disk is truncated at $r_{\rm tr} = 225$, inside of which
there is a RIAF; the accretion rate decreases inwards according to
$\dot{m}(r) = 6.4 \times 10^{-3} \left( r/r_{\rm tr} \right)^{0.8}$ (see text).
The starburst model includes the Fe II emission-line. }
\label{fig:riaf}
\end{figure*}

In Figure \ref{fig:riaf} we show the predicted spectrum of the
accretion flow for $\dot{m}_0=6.4 \times 10^{-3}$ ($\dot{M}_0$ in Eddington units), $p=0.8$ and $\delta=0.1$; the
solid line shows the RIAF contribution and the dashed line
shows the emission of the truncated thin disk.
It is encouraging that we obtain a good fit of the optical and
X-ray portion of the SED using the value of $r_{\rm tr} = 225$ derived
from the line profile model, i.e.  our models for the SED and the line
profile are consistent with each other!

As shown by \citet{quataert99b}, when comparing spectral models of
RIAFs with observations, there are degeneracies between the mass-loss
rate in the wind (parameter $p$) and parameters describing the
microphysics of the RIAF, particularly $\delta$, so the fits to the
data presented here are not unique; instead they should be treated as
representative models. For example, we can fit the SED equally well with an ADAF model, where $p=0$ and $\delta=0.01$.
It is worth noting that there are theoretical uncertainties on the value of $\delta$ \citep{quataert99c}, and that as the
accretion rate increases, it is expected that outflows/convection
become less well developed, because the Bernoulli parameter decreases
with increasing $\dot{m}$.

The accretion flow model severely underpredicts
the radio emission and underpredicts to a small degree the higher
frequency UV data.
Underprediction of radio data is common in RIAF models
\citep[e.g.,][]{quataert99,ptak04,ulvestad01,anderson04,wu05}, and the
radio emission would be enhanced if the source has a jet (e.g.,
\citealt{yuan02a,yuan02b,yuan05}, hereafter YCN05) or if the RIAF has
nonthermal electrons \citep{mahadevan99,ozel00,yuan03}.  To test the
possibility that the observed radio emission comes from a jet, we use
the model for the jet emission of YCN05, which adopts the internal
shock scenario widely employed in the interpretation of gamma-ray
burst afterglows. Since the only constraint on the jet emission is the
observed radio flux at 8.4 GHz, we assume that the most basic
parameters have the same values as in YCN05. Namely the half-opening
angle is $\phi = 0.1$, the energy density of accelerated electrons is
$\epsilon_e = 0.06$, and the amplified magnetic field energy
density is $\epsilon_B = 0.02$. The bulk Lorentz factor of the jet is
$\Gamma_j = 10$, the viewing angle is $34^{\circ}$ (same as the
inclination angle of the thin accretion disk inferred by SB03) and the
mass-loss rate in the jet is $\dot{M}_{\rm jet} = 7 \times 10^{-7}
\dot{M}_{\rm Edd}$, which is about 0.5\% of the mass accretion rate of
the ADIOS at 5 $R_S$. The resulting jet spectrum is shown in Figure
\ref{fig:riaf} as the dotted line. We see that this jet model accounts
for the available high-resolution radio data. The radio band is
dominated by the jet, and the jet emission becomes less important in
the rest of the SED compared to the accretion flow emission.  We note
that as we have so few constraints on the jet emission, the jet
parameters are not as well constrained as those of the accretion flow
models, although the results are not very dependent on the values of
the jet parameters.

The discrepancy between the model and the observed UV flux can be
resolved by invoking the contribution of a compact starburst to the
observed UV spectrum, as found by SB05. The starburst has a mass of
approximately 10$^6$ solar masses, an age of at most a few $\times$
10$^6$ yrs and is obscured by $A_V \approx 3 \, {\rm mag}$. The
signatures of the starburst include a number of absorption features in
the 1000--1600 \AA\ spectral region and the continuum contribution
illustrated in Figure \ref{fig:riaf} as the dot-dashed line.  Included
in the theoretical SED is an Fe~II template broadened according to the
width of the Balmer lines, which accounts for the observed bump around
2500 \AA\ (see SB05 for details). This UV Fe~II emission presumably
comes from the thin disk.

We note that while the starburst is reddened by $A_V=3$  mag, the X-ray data, the double-peaked lines (including the UV Mg II lines) and our successful modelling suggest that the AGN is unobscured, supporting the suggestion by SB05 that the starburst could be in a circumnuclear dusty structure, which does not intercept our line of sight to the nucleus.

We also notice (see Figure \ref{fig:riaf}) that our best model somewhat overpredicts the luminosity of the UV end of the HST spectrum. This part of the spectrum is very sensitive to the amount of dust extinction. If the value of $A_V$ is slightly larger, then the intrinsic UV flux would be higher than the adopted one and would agree better with the model. Interestingly enough, \citet{prieto05} measured $A_V \sim 1$ within the central $\sim 60 - 40$ pc region, a higher extinction than the value adopted in this work.
We can obtain a small improvement in the fit of the UV end varying the parameters $\dot{m}_0$ and $p$ of the RIAF model, but in that case the quality of the fit at the red end of the HST spectrum will worsen.

\section{DISCUSSION} \label{sec:discuss}

The adopted transition radius ($r_{\rm tr} \approx 225$) is similar to
the typical values found in available RIAF + thin disk models of LINER
SEDs (NGC 4579 and M81: $r_{\rm tr} \approx 100$, \citealt{quataert99};
NGC 4258: $r_{\rm tr} \approx 10-100$, \citealt{gammie99}; NGC 3998:
$r_{\rm tr} = 300$, \citealt{ptak04}).

We tried to fit the OUV data with a model of a thin disk extending down to the ISCO at 3 $R_{\rm S}$.
Although a model with $\dot{m}=3 \times 10^{-5}$ fits the UV data, it cannot account for lower frequency optical data and it would imply that $\approx 50\%$ of
$L_{\rm bol}$ is due to X-rays (possibly emitted by a corona above the
disk), something very unusual for an AGN (e.g.,
\citealt{haardt93}). This fit shows that truncation of the thin disk
is definitely needed in order to obtain a consistent picture for the
nature of the low optical flux, as found in similar models of other LINERs (e.g., \citealt{quataert99,ptak04}).

\citet{yuan04} (hereafter YN04) assembled a plot showing the
dependence of $L_{\rm bol} / L_{\rm Edd}$ on $r_{\rm tr}$ for many
accreting black hole sources fitted with RIAF models, including for
example Sgr A$^*$, X-ray binaries and LLAGNs (Figure 3 of YN04). This
plot relates two fundamental properties of accreting black holes and
provides a way of putting in context these sources. Interestingly
enough, when the corresponding values of \ngc\ are inserted in this
plot, we find that \ngc\ falls within the relation found by YN04, and
is located very close to other LINERs previously modeled with RIAFs
(M81 and NGC 4579, \citealt{quataert99}).

Using the estimated accretion rate from the accretion flow
model and the black hole mass, we can make an estimate of the gravitational power output of
the thin disk as $W_{\rm disk} = 2.1 \times 10^{42} \; \rm{erg} \;
\rm{s}^{-1}$, which is higher than the
H$\alpha$ luminosity by two orders of magnitude. Although this calculation suggests that in the
particular case of \ngc\ the energy budget of the thin disk is such
that the disk could power locally its own emission lines, the direct
conversion of gravitational energy into line emission is improbable,
because most of the binding energy of the thin disk is radiated in the
form of a weakly ionizing IR/optical bump which is not capable of
photoionizing the atmosphere of the thin disk and therefore cannot
contribute to the formation of the broad double-peaked emission
lines. Furthermore, there are
theoretical uncertainties in the accretion flow model which
affect the determination of $\dot{m}$.

Current observations do not allow us to distinguish between the
different RIAF models, although there are strong theoretical
reasons favoring ADIOS / CDAF (e.g., \citealt{quataert01}) over to ADAF models. There are some future
observations that may help us assess the importance of outflows in
RIAFs. For instance, the absence of linear polarization
in the high-frequency radio spectrum of \ngc\ would argue in favor of
ADIOS/CDAF models, even in the presence of a jet
\citep{quataert00rad}. This would be attainable with high-frequency
VLBI interferometry.
More radio observations to constrain the shape of the radio spectrum
are also needed to verify whether it is consistent with the predictions
of the assumed jet model.

\section{SUMMARY AND CONCLUSIONS} \label{sec:conc}

We have reproduced the nuclear optical to X-ray SED of \ngc, composed
of HST and \emph{Chandra} observations, using a model in which the
accretion flow consists of an optically thin, geometrically thick RIAF
for radii smaller than the transition radius $r_{\rm tr} = 225 R_S$
and an optically thick, geometrically thin disk for larger
radii. The value of $r_{\rm tr}$ is the same as the inner disk
radius obtained from fitting the double-peaked H$\alpha$ line profile
(SB03) and the value of the black hole mass is that inferred from the
stellar velocity dispersion of the host galaxy. Thus, an important
result of this work is the consistency between the result of fitting
the SED with the accretion flow model and the double-peaked Balmer
line with the thin disk model, i.e. we obtain acceptable fits of these
two different data sets with the same transition radius.

The accretion flow model underpredicts slightly the UV continuum, but
the added contribution of a young and obscured starburst of mass $\sim
10^6 \; M_\odot$ (SB05) can account for the excess UV emission.  The
accretion model also underpredicts the radio emission, suggesting the
need for a jet component. We have been able to reproduce the radio
data with a jet model employing the internal shock scenario (YCN05),
although the jet parameters are not as well constrained as those of
the accretion flow models, because we have few constraints on the jet
emission.

The results of this work support the proposal that the basic
components of the nuclei of LLAGNs are a RIAF, a truncated thin disk
and a jet (e.g., \citealt{ho05}), with the possibly important role of
starbursts presumably obscured by the circumnuclear torus predicted in
the unified model of AGNs.




\acknowledgments

The authors are grateful to an anonymous referee for a
number of suggestions which helped improve the presentation.
RSN thanks the hospitality of ICC at Durham University, where part of this work was written, and useful discussions with Hor\'acio
Dottori, Jo\~ao Steiner and Luiz Fernando Ziebell.
FY acknowledges support from the One-Hundred-Talent Program of China.
This work was
supported by the Brazilian institutions CNPq, CAPES and FAPERGS, and
in part by NASA through grants NAG81027 and NAG5-13065 (LTSA) to the
University of Maryland (ASW) and through grants NAG5-10817 (LTSA) and
HST-GO-08684.01-A (from the Space Telescope Science Institute, which
is operated by AURA, Inc. under NASA contract NAS5-26555) to the
Pennsylvania State University (ME).
This research has made use of the
NASA/IPAC Extragalactic Database (NED) which is operated by the Jet
Propulsion Laboratory, California Institute of Technology, under
contract with the National Aeronautics and Space Administration.




\clearpage
\appendix
\section{X-RAY SPECTRA OF CIRCUMNUCLEAR STAR-FORMING REGION AND
ULTRALUMINOUS COMPACT X-RAY SOURCE} \label{ap:ring_ulx}

X-ray spectra of the circumnuclear star-forming region at $\sim
9\arcsec$ from the nucleus\footnote{Note that this is not the nuclear starburst found by SB05} and the ULX
were taken with the \textit{Chandra X-ray Observatory}. These were
analyzed as described in the main text, with the following results.

\begin{figure}[!ht]
\centering
\includegraphics[scale=0.35,angle=-90]{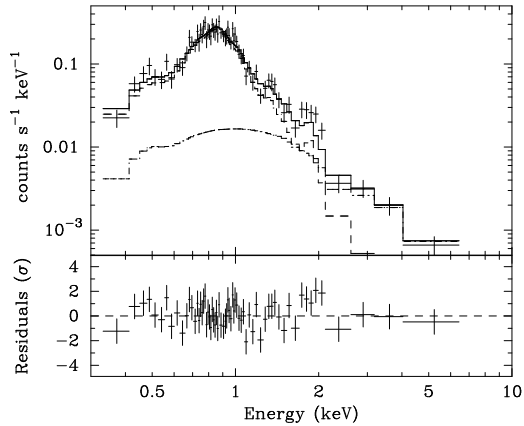}
\caption{X-ray spectrum of the circumnuclear starforming region of
\ngc. The lower dot-dashed curve represents the thermal bremsstrahlung
component, the upper dashed one represents the MEKAL plasma component
and the solid line is the composite model. The lower panel shows the
residuals of the fit of the composite model.}
\label{fig:chandra-ring}
\end{figure}

Figure \ref{fig:chandra-ring} shows the spectrum of the circumnuclear
star-forming region. Since the emission is diffuse, the effect of pile
up is negligible for both observations. We performed spectral fits to
the combined spectrum of the two observations. We applied a MEKAL
plasma model to the spectrum, but the resulting fit was unacceptable,
with significant positive residuals remaining above 2 keV. For this
fit, we obtained $N_{\rm H} = 1.2 \times 10^{21} \: \textrm{cm}^{-2}$,
$k T = 0.63 \; \textrm{keV}$, and an abundance of $0.099$ times solar
(with $\chi^2 = 114.0$ for 62 dof).

Since the simple MEKAL model fit was poor, we added a hard component
to the model, represented by a thermal bremsstrahlung model with $kT
= 7 \; \textrm{keV}$. This model has been successfully applied in the
case of superposition of emission from low-mass X-ray binaries
\citep{makishima89} and results in a better fit to the spectrum, as
illustrated in Figure \ref{fig:chandra-ring}. The parameters for the
spectral fit are listed in Table \ref{tab:chandra-fits}.

\begin{figure}[!ht]
\centering
\includegraphics[scale=0.35,angle=-90]{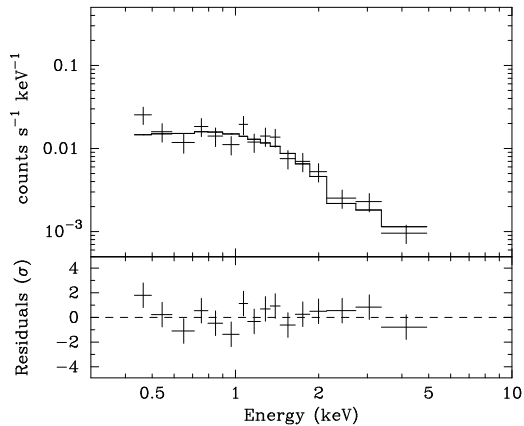}
\caption{X-ray spectrum of the ultraluminous X-ray source (ULX) of
\ngc. The solid line represents the power-law model.}
\label{fig:chandra-ulx}
\end{figure}

The ULX has a spectral shape and flux in the two observations which
are consistent with each other within statistical errors. Figure
\ref{fig:chandra-ulx} shows the spectrum of the ULX. Spectral fits
were performed to the combined spectrum of the two observations using
a maximum-likelihood method. A
power law model gives a better fit than a multi-color disk blackbody
model (MCD). The resulting parameters for the power law model are
listed in Table \ref{tab:chandra-fits}, assuming the object is
associated with \ngc.

\clearpage

\begin{deluxetable}{ccccc}
\tabletypesize{\scriptsize}
\tablecaption{X-ray observation log. \label{tab:chandra-log}}
\tablewidth{0pt}
\tablehead{
& & & & \colhead{Count Rate} \\
\colhead{Observation ID} & \colhead{Exposure (s)\tablenotemark{a}} &
\colhead{CCD Mode} & \colhead{Nuclear Counts\tablenotemark{b}} &
\colhead{per Second}
}
\startdata
1611    &   5343    &   Full array  &   755 &   0.14\\
2339    &   5442    &   1/8 sub-array   &   1690    &   0.31\\
\enddata
\tablenotetext{a}{Effective exposure time after data screening.}
\tablenotetext{b}{Counts obtained in the nuclear region
($r=2\farcs6$), not corrected for pile up. The number of counts in the
first observation is small because of significant pile up.}
\end{deluxetable}

\begin{deluxetable}{ccccc}
\tabletypesize{\scriptsize}
\tablecaption{Power-law fits to X-ray spectra \label{tab:chandra-fits}}
\tablewidth{0pt}
\tablehead{
& $N_{\rm H}$ & & $F_{2-10 \: \rm{keV}}$ & \colhead{} \\
\colhead{Component} & \colhead{($\times 10^{20} \: \textrm{cm}^{-2}$)} &
\colhead{$\Gamma$} & \colhead{($10^{-12} \: \rm{erg} \; {s}^{-1} \:
\rm{cm}^{-2}$)} & \colhead{$\chi^2/\rm{dof}$}
}
\startdata
Nucleus\tablenotemark{a} & $2.3^{+2.8}_{-1.7}$ &
$1.64^{+0.13}_{-0.07}$ & 1.73 &  60.5/70\\
Star-forming ring\tablenotemark{b} &   $7.7^{+3.1}_{-3.0}$ & -- &
0.151\tablenotemark{c} & 63.4/61\\
ULX & $0^{+3.8}_{-0}$ & $1.55^{+0.37}_{-0.18}$ & 0.14 & 11.9/13\\
\enddata
\tablenotetext{a}{Corrected for pileup. The normalization constant is
$3.88^{+0.39}_{-0.1} \times 10^{-4} \; \rm{photons} \;  \rm{keV}^{-1}
\; \rm{cm}^{-2} \; \rm{s}^{-1}$ at 1 keV (after correction for
absorption). The errors quoted represent the 90\% confidence level for
one parameter of interest ($\Delta \chi^2 = 2.7$).}
\tablenotetext{b}{MEKAL plasma + thermal bremsstrahlung model. The best-fit parameters for the MEKAL component are $kT = 0.61 ^{+0.03}_{-0.04} \; \textrm{keV}$ and an abundance of $0.18
^{+0.08}_{-0.05} \; \textrm{solar}$. $kT$ = 7 keV is assumed for the thermal bremsstrahlung component.}
\tablenotetext{c}{Bremsstrahlung component only. The MEKAL component
contributes an observed flux in the 0.5 -- 4 keV band of
  $5.04 \times 10^{-13} \: \textrm{erg s}^{-1} \textrm{cm}^{-2}$
(corrected for absorption).}
\end{deluxetable}

\begin{deluxetable}{cccccc}
\tabletypesize{\scriptsize}
\tablecaption{Radio and infrared data for the nucleus of \ngc.
\label{tab:ir-radio}}
\tablewidth{0pt}
\tablehead{
\colhead{$\nu$ (Hz)} & \colhead{$S_\nu$ (Jy)} & \colhead{Uncertainty
(Jy)} & \colhead{$\nu L_\nu$ (erg s$^{-1}$)} & \colhead{Resolution (")}
& \colhead{Reference}
}
\startdata
$8.4 \times 10^{9}$ &  0.0031 & $4.51 \times 10^{-4}$ & $1.24 \times
10^{36}$ & 0.25 & 1 \\
$1.5 \times 10^{13}$ &  0.240 & 0.073 & $8.89 \times 10^{41}$ & 5 & 2 \\
$2.94 \times 10^{13}$ & 0.065 & 0.009 & $4.67 \times 10^{41}$ & 5 & 2 \\
$8.47 \times 10^{13}$ &  0.0731 &    -- & $8.71 \times 10^{41}$ & 6 & 3 \\
$1.35 \times 10^{14}$ & 0.0409 &    -- & $2.48 \times 10^{42}$ & 6 & 3 \\
$1.39 \times 10^{14}$ & 0.0802 &    -- & $2.80 \times 10^{42}$ & $0.18 \times 0.15$\tablenotemark{a} & 4 \\ 
$1.87 \times 10^{14}$ & 0.0878 &    -- & $4.14 \times 10^{42}$ & 6 &
3 \\
$1.80 \times 10^{14}$ & 0.049 &    -- & $2.22 \times 10^{42}$ & $0.20 \times 0.18$\tablenotemark{a} & 4 \\ 
$2.40 \times 10^{14}$ & 0.0604 &    -- & $3.64 \times 10^{42}$ & 6 &
3 \\
$2.43 \times 10^{14}$ & 0.023 &    -- & $1.41 \times 10^{42}$ & $0.20 \times 0.19$\tablenotemark{a} & 4 \\ 
\enddata
\tablerefs{
(1) \citealt{thean00}
(2) \citealt{telesco81}
(3) \citealt{glass85}
(4) \citealt{prieto05}
}
\tablenotetext{a}{The quoted aperture is the FWHM of the achieved spatial resolution in the respective band. }
\end{deluxetable}

\clearpage

\end{document}